\documentclass[12pt]{iopart}
\usepackage[latin1]{inputenc}

\usepackage{iopams}

\newcommand{\kF}{k_\mathrm{F}}
\newcommand{\nag}{{\phantom{\dag}}}

\usepackage{color}
\usepackage{graphicx}
\begin{document}

\title[Peierls to superfluid crossover in the 1D Holstein model]{Peierls to superfluid crossover in the one-dimensional, quarter-filled Holstein model}

\author{M Hohenadler and F F Assaad}

\address{Institut f\"ur Theoretische Physik und Astrophysik, Universit\"at W\"urzburg,\\ Am Hubland, 97074 W\"urzburg, Germany}

\begin{abstract}
  We use continuous-time quantum Monte Carlo simulations to study retardation
  effects in the metallic, quarter-filled Holstein model in one
  dimension. Based on results which include the one- and two-particle spectral functions as
  well as the optical conductivity, we conclude    that with increasing
  phonon frequency the ground state evolves from one
  with dominant diagonal order---$2\kF$ charge correlations---to one
  with dominant off-diagonal fluctuations, namely s-wave pairing correlations. In the
  parameter range of this crossover, our numerical results support the
  existence of a spin gap for all phonon frequencies. The crossover can hence
  be interpreted in terms  of preformed pairs corresponding to bipolarons,
  which are essentially  localised in the Peierls phase, and ``condense''
  with increasing  phonon frequency to generate dominant pairing correlations.  
\end{abstract}

\section{Introduction}

The concept of preformed fermion pairs or bosonic degrees of freedom which
condense to form a superfluid can be found in many domains of correlated
quantum many-body systems. Examples include the resonating valence bond theory
of high-temperature superconductivity \cite{Anderson04}, Mott metal-insulator
transitions in cold atoms \cite{Greiner00}, or transitions between
charge-density-wave and superconducting states in the family of
dichalcogenides \cite{Sipos08}.  In this article, we show in the framework of
the Holstein model that these concepts can be carried over to one dimension
where,  in the absence of continuous symmetry breaking, the phase transition
is replaced by a crossover.

Quantum lattice fluctuations play a crucial role in one dimension. 
For classical phonons (i.e., in the adiabatic limit), the Peierls
instability leads to a $2\kF$ periodic lattice deformation, resulting in
corresponding charge ordering of electron pairs (bipolarons) and the opening of a gap
at the Fermi level. In contrast, for the Holstein model, one can argue and
give numerical evidence \cite{JeZhWh99} that quantum phonons will
close the charge gap but preserve the spin gap, thereby giving rise to a
``bad metal'' state.  In the latter, electrons bind into pairs and are mostly
localised in the $2\kF$ charge order pattern. This results in a very high
effective charge-carrier mass or, equivalently, in a very small Drude weight. We refer to this state as 
``the Peierls state'', which is to be distinguished from an insulating Peierls
state with long-range charge order \cite{Peierls}. The latter has been
studied intensely in Holstein-type models with and without spin, see discussion in
section~\ref{sec:method}. The metallic Peierls state may be regarded as a
more general concept for quasi-one-dimensional systems, applicable to
noncommensurate fillings or the normal state ($T>T_\mathrm{c}$) of a Peierls
insulator \cite{Voit98}. On the level of bosonization, a metallic state with
dominant $2\kF$ charge correlations requires a finite spin gap \cite{Voit98}.
In the anti-adiabatic limit, the lattice adjusts instantaneously to the
electronic motion, and the Holstein model maps onto the attractive Hubbard
model with dominant pairing correlations.  

The question addressed in this article is the nature of the crossover from
the adiabatic to the anti-adiabatic regime. The limiting cases of low (but
finite) and high (or infinite) phonon frequency can be described as a
Luther-Emery liquid \cite{Emery79,Giamarchi}, corresponding to a Luttinger
liquid with a gapless charge mode and a gapped spin mode. A key result of the
present work is that the crossover occurs without a closing of the spin gap,
which suggests that the bipolarons existing in the Peierls state do not
dissociate as a function of phonon frequency but instead constitute the
preformed pairs of the superfluid phase. 

The article is organised as follows.  In  section~\ref{sec:method} we 
introduce the model and explain the key ideas of the continuous-time quantum
Monte Carlo method. Numerical results are presented in
section~\ref{sec:results}, and we end with conclusions and a discussion of
open issues and future research directions in section~\ref{sec:discussion}.

\section{Model and method}\label{sec:method}

The one-dimensional Holstein model is defined by the Hamiltonian \cite{Ho59a}
\begin{eqnarray}
\label{Holstein_Ham}
  \hat{H}  
  &&=
  \sum_{k\sigma} \epsilon^\nag_k \hat{c}^{\dag}_{k\sigma} \hat{c}^\nag_{k\sigma}  
  + \sum_{i} \left(\mbox{$\frac{1}{2M}$} \hat{P}_{i}^2 + \mbox{$\frac{K}{2}$}
    \hat{Q}_{i}^2 \right)\\\nonumber
  &&\quad- g \sum_{i} \hat{Q}_{i} \left( \hat{n}_{i}-1\right) 
\,,
\end{eqnarray}
with the tight-binding dispersion relation $ \epsilon_k = -2t \cos(k a)-\mu$
and the chemical potential $\mu$. The operator $ \hat{c}^{\dag}_{i\sigma} $
creates an electron in the Wannier state centred on lattice site $i$ with spin
$\sigma$, $\hat{c}^{\dag}_{k\sigma } = L^{-1/2} \sum_{j} \rme^{ \rmi k j }
\hat{c}^{\dag}_{j\sigma}$ creates an electron in a Bloch state with
wavevector $k$ and spin $\sigma$, and $ \hat{n}_{i} = \sum_{\sigma} \hat{c}^{\dag}_{i\sigma
}\hat{c}^\nag_{i\sigma } $ is the particle number operator; $\hat{Q}_{i} $
and ${\hat P}_{i} $ denote the displacement and momentum of the harmonic
oscillator at site $i$.  

To solve this model numerically without approximations (except for the finite
system size), we integrate out the phonons at the expense of a retarded
density-density interaction. The partition function of the resulting model is
given by the path integral
\begin{equation}	
  Z = \int  \left[\rmd c^{\dag} \rmd c\right] \rme^{- ( S_0 + S_{\rm ep})}  \nonumber
\end{equation}
with the following contributions to the action:
\begin{eqnarray}\label{eq:S0}
  S_0 && =  \int_{0}^{\beta} \rmd \tau    \sum_{i,j,\sigma} c^{\dag}_{i\sigma}(\tau) 
  \left( \delta_{ij} \partial_\tau - t_{ij} \right)  c^\nag_{j\sigma}(\tau)\,,
\end{eqnarray}
where $t_{ij}=1$ for nearest neighbours and zero else, and
\begin{eqnarray}\label{eq:S1}
  S_{\rm ep} &&=\int_{0}^{\beta} \rmd \tau \int_{0}^{\beta} \rmd \tau'
  \sum_{i,j} 
  \left[n_i(\tau )-1\right] D_0(i-j,\tau-\tau')
                          \left[n_j(\tau')-1\right]\,.
\end{eqnarray}
Here, $c^{\dag}_{i\sigma}(\tau)$ is a Grassmann variable, $\beta$ denotes
the inverse temperature, and  $D_0(i-j,\tau-\tau')$ is the phonon propagator.
For dispersionless Einstein modes, $D_0$ is diagonal in space and takes the form
\begin{eqnarray}\label{eq:prop}
  &D_0(i-j,\tau - \tau') 
  = 
  \delta_{ij}\frac{g^2}{2 k} P( \tau - \tau')
  \,,\\\nonumber
  &P( \tau )  
  =
  \frac { \omega_0 } { 2} 
  \frac{\rme^{-|\tau|\omega_0} + \rme^{ - ( \beta - | \tau| ) \omega_0 }}{ 1-\rme^{-\beta \omega_0}}\,.
\end{eqnarray}
The phonon frequency is given by $\omega_0 = \sqrt{{K}/{M}} $. In the
following, we use $t$ as the unit of energy, and set $\hbar$ and the
lattice constant to unity.

In the fermionic form given by equations~(\ref{eq:S0}) and~(\ref{eq:S1}), the
problem can be solved very efficiently by means of interaction-expansion continuous-time quantum Monte Carlo methods
\cite{Rombouts99,Rubtsov05,Gull_rev}.  The advantage of such a formulation is
that we do not have to sample the lattice degrees of freedom directly,
thereby eliminating the need for a Hilbert space cutoff. The details of the implementation of
this algorithm can be found in \cite{Assaad07}. The CTQMC method has
previously been applied to the spinless \cite{Hohenadler10a} and the spinful
Holstein model \cite{Assaad08}, as well as to a model with nonlocal
electron-phonon interaction \cite{Ho.As.Fe.12}. Simulations are carried out
on one-dimensional lattices of length $L$ using periodic boundary conditions.

The one-dimensional, half-filled Holstein model---with or without an
additional Hubbard term to describe electron-electron repulsion---has been
studied for many years. The existence of a metal-insulator transition at
zero temperature has been established with the help of exact numerical methods,
including quantum Monte Carlo
\cite{Hirsch83a,ClHa05,hardikar:245103,PhysRevB.84.165123,Ho.As.Fe.12} the
density-matrix renormalization group
\cite{JeZhWh99,ZhJeWh99,0295-5075-84-5-57001,1742-6596-200-1-012031,TeArAo05,MaToMa05}
and exact diagonalization \cite{FeWeHaWeBi03}. An overview of analytical
work on the Peierls transition can be found in \cite{Ba.Bo.07}.
Many aspects of the relevant physics can also be captured by a spinless model with filling $n=0.5$,
including an extended metallic region in the phase diagram
\cite{BuMKHa98,WeFe98,Hohenadler06,Ej.Fe.09}, the renormalization of the phonon mode
\cite{Hohenadler06,CrSaCa05,SyHuBeWeFe04,Sykora06}, and soliton excitations
\cite{Hohenadler10a}. Retardation effects in the half-filled Holstein model were
discussed, for example, in \cite{tam:161103,Fehske00}, and for
the spin-Peierls transition in \cite{Ci.Or.Gi.05}.

Here we study the effect of phonon frequency in the {\it quarter-filled} Holstein
model at a fixed electron-phonon coupling strength. For quarter filling and low
phonon frequencies, the absence of first-order umklapp scattering---dominant at half
filling---suppresses the Peierls instability, and stabilises a state with
gapless charge but gapped spin degrees of freedom, as well as a nonzero
single-particle excitation gap, and dominant but critical $2\kF$ charge
correlations (a Luther-Emery liquid with interaction parameter $K_\rho<1$).  The spin gap is a result of
backscattering. Signatures of this Luther-Emery phase have also been observed in the
Holstein-Hubbard model with comparable electron-phonon and electron-electron
interactions \cite{PhysRevB.67.081102,ClHa05,hardikar:245103,0295-5075-84-5-57001,1742-6596-200-1-012031}.
Our results presented below suggest that in contrast to half filling, we can
observe dominant pairing correlations for high phonon frequencies at a
commensurate density; dominant pairing for incommensurate filling has been
reported before \cite{PhysRevB.76.155114}. Recently,  the existence of
a correlated singlet state in the Holstein-Hubbard model away from half
filling has been suggested \cite{PhysRevB.84.085127,Re.Ya.Li.11}. Compared to
half filling, the critical value for the transition to a Peierls insulator
is significantly larger at quarter filling \cite{hardikar:245103}. This increase
may be understood in terms of a devil's staircase, similar to
charge-density-wave transitions in extended Hubbard models away from half
filling \cite{Giamarchi}. As for half filling, the critical value of the
electron-phonon coupling strength for the metal-insulator transition
increases with increasing phonon frequency as a result of enhanced lattice
fluctuations \cite{PhysRevB.76.155114}.

\section{Results}\label{sec:results} 

In the anti-adiabatic limit $\omega_0\to\infty$, $ P( \tau ) $ in
equation~(\ref{eq:prop}) reduces to a Dirac
$\delta$-function, facilitating the above-mentioned mapping of the Holstein
model onto the attractive Hubbard model, with $U ={g^2}/{ k} $. The ratio of
this  binding energy and the bandwidth $ W = 4t$ gives the dimensionless
electron-phonon coupling
\begin{equation}
  \lambda  =   \frac{g^2}{k W}. 
\end{equation}
Throughout this article, we set $\lambda = 0.35 $ and concentrate on the
quarter-filled band with $\kF = \pi/4 $. In the following, we first establish
a picture of the physics on the basis of equal-time correlation functions,
and then turn to dynamical correlations.

\subsection{Static correlation functions}

\begin{figure}[t]
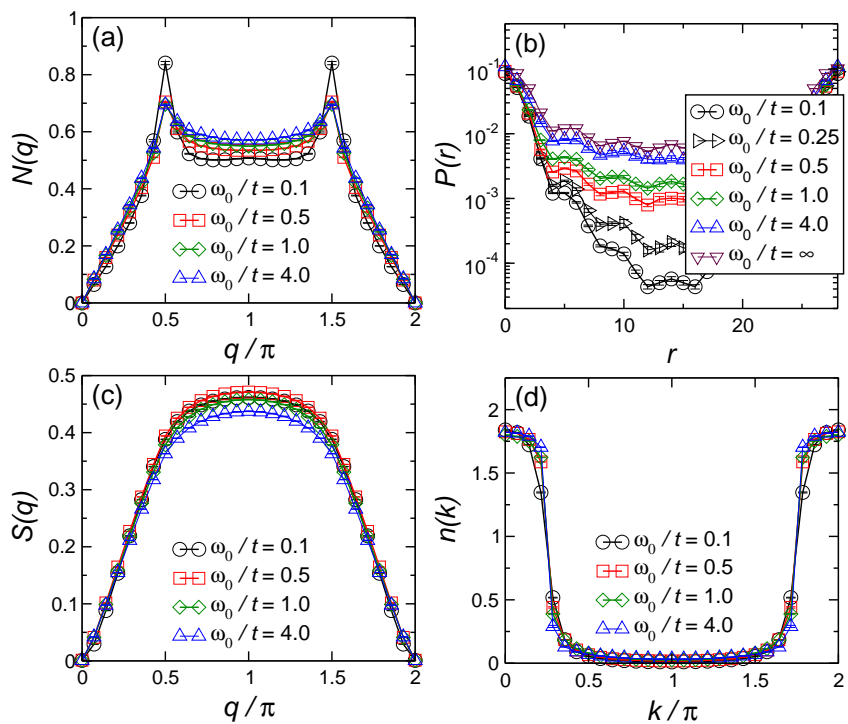

\begin{center}
\includegraphics*[width=0.35\textwidth]{DenJ_om} 
\includegraphics*[width=0.35\textwidth]{PairJ_om} 
\\
\includegraphics*[width=0.35\textwidth]{SpinJ_om} 
\includegraphics*[width=0.35\textwidth]{XnkJ_om} 
\end{center}
\caption{Static correlation functions for different values of the phonon
frequency $\omega_0/t$ at $\lambda=0.35$, and for a quarter-filled band ($n=0.5$).
The panels show (a) the charge structure factor, (b) the pairing correlator,
(c) the spin structure factor, and (d) the momentum distribution function.
Here $L=28$ and $\beta t=40$.} 
\label{Static.fig}
\end{figure}

Figure~\ref{Static.fig}  shows equal-time correlation functions for charge,
spin, and pairing, as well as the momentum distribution function, at various
phonon frequencies. In the adiabatic limit $\omega_0/t=0$, any $\lambda>0$ leads
to an insulating state. As discussed in section~\ref{sec:method}, we choose the
coupling strength $\lambda=0.35$ such that we have a metallic Luther-Emery liquid 
with dominant $2\kF$ charge correlations for low phonon frequencies, and
then study the evolution as a function of increasing $\omega_0/t$. In
particular, we have verified that for $\omega_0/t=0.1$, the lowest nonzero
frequency considered in the following, there is no long-range order; this can
be seen from the finite-size dependence of the charge susceptibility \cite{hardikar:245103}.

The density (or charge) structure factor, defined as 
\begin{equation}
      N(q) =  \sum_{r}  \rme^{\rmi q r} \left( \langle \hat{n}_{r}  
      \hat{n}_{0} \rangle -   \langle \hat{n}_{r} \rangle \langle  \hat{n}_{0} \rangle \right),
\end{equation} 
is plotted in figure~\ref{Static.fig}(a).  For classical phonons, $\omega_0
= 0$, the Peierls instability leads to long-range $2\kF$ charge order at
zero temperature.  As discussed above, quantum lattice fluctuations
(occurring for $\omega_0 > 0 $) can melt this order, and lead to a state with
dominant but power-law $2\kF$ charge correlations \cite{Assaad08}, as
confirmed by the cusp at $2\kF = \pi/2$ in figure~\ref{Static.fig}(a).  The
magnitude of the peak at $q=2\kF$ initially decreases and then saturates upon
increasing the phonon frequency, signalling competing ordering mechanisms as
well as enhanced lattice fluctuations. The linear form
of the charge structure factor at long wavelengths [see figure~\ref{Static.fig}(a)]
indicates a $1/r^2$ power-law decay of the real-space charge correlations and
hence a metallic state.

In figure~\ref{Static.fig}(b), we present the pair correlation function in
the onsite s-wave channel,
\begin{equation}
  P(r) = \langle \hat{\Delta}^{\dag}_{r} \hat{\Delta}^\nag_{0}  \rangle\,,\quad
  \hat{\Delta}^{\dag}_{r}  =  \hat{c}^{\dag}_{r\uparrow} \hat{c}^{\dag}_{r\downarrow}\,. 
\end{equation}
In contrast to the density correlator which picks up diagonal order,
$P(r)$ detects off-diagonal order characteristic of a superconducting state.
In the Peierls state obtained for classical phonons, diagonal long-range
charge order leads to an exponential decay of pairing correlations at
long distances.  The fluctuations resulting from a finite phonon frequency
close the charge gap, and render the pairing correlations critical.
Comparing figures~\ref{Static.fig}(a) and~\ref{Static.fig}(b), we see that
the suppression of the $2\kF$ charge correlations is accompanied by an
increase of the pairing correlations, especially at large distances.  A
possible interpretation is that with increasing phonon frequency, the
trapping of bipolarons in the $2\kF$ lattice modulation gives way to a ``{\it
  condensation}'' (in the usual sense of superfluidity in one dimension) of
those preformed pairs.

\begin{figure}[t]
\begin{center}
\includegraphics*[width=0.5\textwidth]{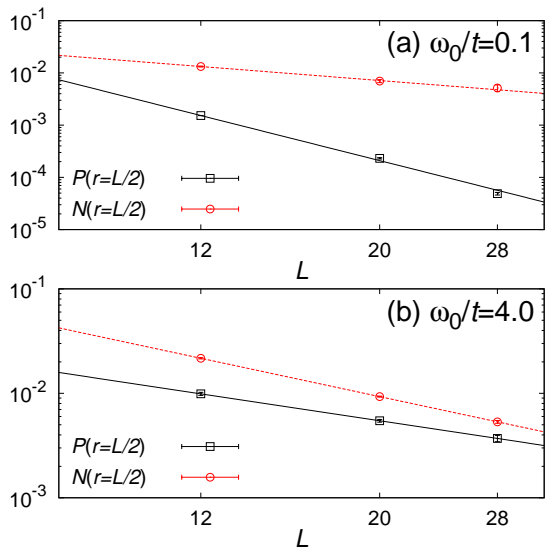} 
\end{center}
\caption{Pairing ($\opensquare$) and charge-density-wave ($\opencircle$)
correlations at the largest distance $r=L/2$ for different system sizes $L$. Here
$\lambda=0.35$, $\beta t=L$ and (a) $\omega_0/t=0.1$, (b)
$\omega_0/t=4$. Lines are fits to the form $f(r) = A/r^\eta$.} 
\label{fig:cdw-pairing}
\end{figure}

The above interpretation relies on the electron pairs remaining bound upon
increasing the phonon frequency. Evidence for the existence of bound pairs
is provided by the equal-time spin-spin correlation function
\begin{equation}
  S(q) =  \sum_{r}  \rme^{\rmi q r} \langle
  \hat{S}^z_{r} \hat{S}^z_{0} \rangle\,,
\end{equation}
which we show in figure~\ref{Static.fig}(c). For all values of the phonon
frequency, the cusps at $q=0$ and $q=2\kF$ are smeared out, thus lending
support to an exponential decay of the spin correlations.  The presence
of a spin gap will be explicitly confirmed by results for the dynamical spin
structure factor below.

Finally, we plot in figure~\ref{Static.fig}(d) the momentum distribution function 
\begin{equation}
  n(k)  = \sum_{\sigma} \langle \hat{c}^{\dag}_{k\sigma} 
  \hat{c}^\nag_{k\sigma}  \rangle\,.
\end{equation}  
Following the above interpretation, we  expect a smooth variation of this
quantity as a function of wavevector $k$. In particular we do not expect
any nonanalytical behaviour at $k=\kF$ or $k=3\kF$.  For classical phonons,
the smooth variation of $n(k)$ stems from the nonzero single-particle gap of
the Peierls state.  In the
anti-adiabatic limit, where the model maps onto the attractive Hubbard
model, this behaviour is equally expected since it is known that the ground
state is a Luther-Emery liquid with gapped single-particle excitations \cite{Giamarchi}.
Since we observe a continuous crossover from low to high frequencies in
figure~\ref{Static.fig}(d), we argue that our data supports the existence of
a single-particle gap for all considered values of $\omega_0$; this is
confirmed in figure~\ref{akw.fig} and discussed in the following subsection.

Our analysis of the equal-time correlation functions in this section suggests
the existence of a  crossover from dominant diagonal (charge-density-wave) to
dominant
off-diagonal (pairing) fluctuations as a function of the phonon frequency.
This evolution is confirmed by the results for $P(r)$ and $N(r)=\langle \hat{n}_{r}  
\hat{n}_{0} \rangle -   \langle \hat{n}_{r} \rangle \langle  \hat{n}_{0}
\rangle$ shown in figure~\ref{fig:cdw-pairing}, where we plot these two correlators
at the largest distance $r=L/2$ for different $L$. Whereas for
$\omega_0/t=0.1$ the charge correlations decay significantly slower than the
pairing correlations, see figure~\ref{fig:cdw-pairing}(a), dominant pairing
correlations can be seen for $\omega_0/t=4$ in figure~\ref{fig:cdw-pairing}(b).
The ratio of the exponents obtained from least-square fits is
$\eta_\mathrm{CDW}/\eta_\mathrm{SS}\approx 0.3$ for $\omega_0/t=0.1$, and 
$\eta_\mathrm{CDW}/\eta_\mathrm{SS}\approx 1.4$ for $\omega_0/t=4$.

\subsection{Dynamical correlation functions}

With the help of a stochastic analytical continuation scheme \cite{Beach04a},
we can extract momentum and frequency dependent spectral functions from 
the imaginary time Green's functions accessible in the quantum Monte Carlo
simulations. Here we consider the single-particle spectral function
\begin{eqnarray}\label{eq:akw}
  A(k,\omega)
  &&=
  \frac{1}{Z}\sum_{n,m}
  (\rme^{-\beta E_n}+\rme^{-\beta E_m}) 
  \\\nonumber
  && \quad \times
  | \langle n| \hat{c}_{k\sigma} | m \rangle|^2 
  \delta(E_n - E_m-\omega)
  \,,
\end{eqnarray}
the dynamical charge  structure factor
\begin{eqnarray}\label{eq:nqw}
  N(q, \omega) &&=  \frac{\pi}{Z} \sum_{n,m} \rme^{-\beta E_m } | \langle n 
  | \hat{n}_{q} | m \rangle|^2 \\\nonumber
  && \quad \times \delta( E_n -  E_m - \omega )   \,,
\end{eqnarray}
where $ \hat{n}_{q} = L^{-1/2} \sum_{j} \rme^{\rmi q j} \hat{n}_{j} $
and  with  the sum rule  $ N(q) = \pi^{-1} \int\rmd \omega  N(q,\omega)$, and
the dynamical spin structure factor $S(q, \omega)$ defined as in equation~(\ref{eq:nqw})
but with $\hat{n}_q$ replaced by $\hat{S}^z_{q}$. 

\begin{figure}[t]
\begin{center}
\includegraphics[width=0.5\textwidth]{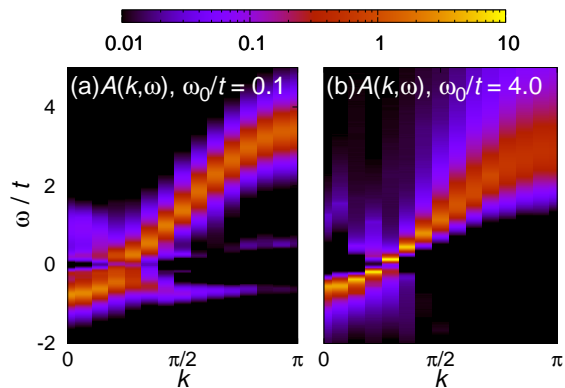}
\end{center}
\caption{Single-particle spectral function for (a)
  $\omega_0/t = 0.1$  and (b) $\omega_0/t = 4$. Here
  $\lambda=0.35$, $n=0.5$, $L=28$, and $\beta t =28$.}
\label{akw.fig}
\end{figure}

The results for $A(k,\omega)$ shown in figure~\ref{akw.fig} reveal the
existence of a spin gap both at low [(a), $\omega_0/t=0.1$] and
high [(b), $\omega_0/t=4$] phonon frequencies, although the size of the gap
is smaller in figure~\ref{akw.fig}(b) than in  figure~\ref{akw.fig}(a). Very
similar results have previously been obtained for $\omega_0/t = 0.1$ \cite{Assaad08}.

\begin{figure}[t]
\begin{center}
\includegraphics[width=0.5\textwidth]{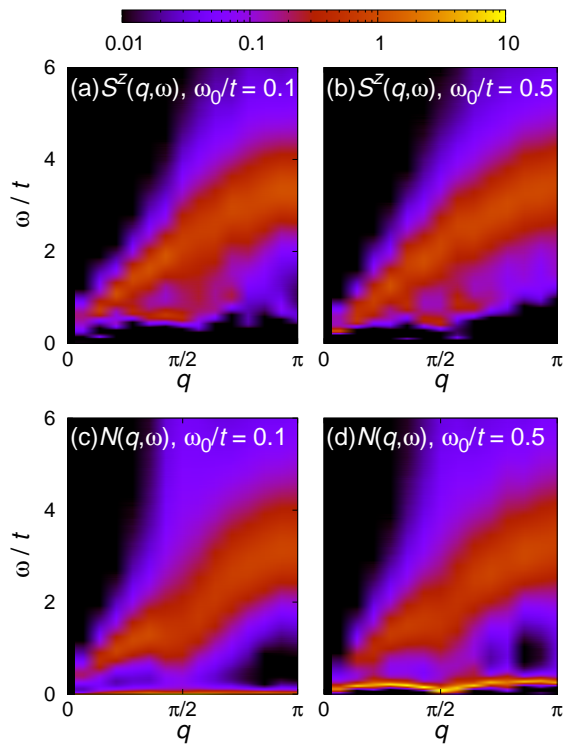}
\end{center}
\caption{Dynamical spin (a,b) and charge (c,d) structure factors for (a,c)
  $\omega_0/t = 0.1$  and (b,d) $\omega_0/t = 0.5$. Here
  $\lambda=0.35$, $n=0.5$, $L=28$, and $\beta t =40$.}
\label{Dynamic.fig}
\end{figure}

The nonzero spin gap is confirmed by our results for the dynamical spin
structure factor $S(q,\omega)$ in figure~\ref{Dynamic.fig}(a) (for
$\omega_0/t=0.1$) and figure~\ref{Dynamic.fig}(b) (for $\omega_0/t=0.5$),
which reveal the absence of low-lying spectral weight near $q=0$.

The suggested crossover from a Peierls to a superconducting state is expected
to be reflected in an enhancement of the mobility of the preformed pairs.
For the two values of the phonon frequency considered here ($\omega_0/t=0.1,
0.5$), the dynamical charge structure factor in figures~\ref{Dynamic.fig}(c)
and (d) shows two features: a high-energy particle-hole continuum reminiscent
of the noninteracting problem, and a low-lying excitation band.  At
$\omega_0/t  = 0.1$, figure~\ref{Dynamic.fig}(c), the dispersion of the
latter is extremely flat and within our resolution we cannot distinguish the
associated charge velocity from zero. This narrow mode reflects the very slow
dynamics of bipolarons which are predominantly localised to form the $2\kF$
charge-density-wave order. 
At a higher phonon frequency, $\omega_0/t=0.5$, see figure~\ref{Dynamic.fig}(d),
a clear dispersion relation  with a minimum at $q=2\kF$ emerges.  

\begin{figure}[t]
  \begin{center}
    \includegraphics*[width=0.45\textwidth]{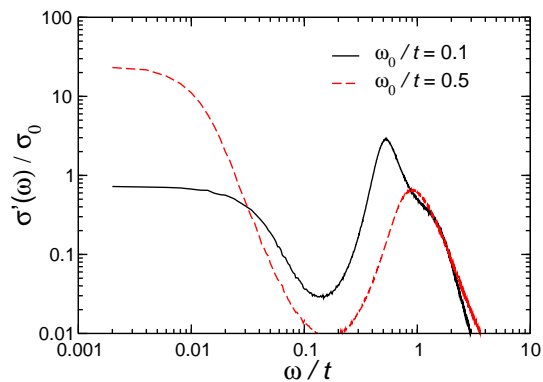} 
  \end{center}
  \caption{Optical conductivity for $\omega_0/t=0.1$ ($\full$) and $\omega_0/t=0.5$ ($\dashed$). Here
    $\lambda=0.35$, $n=0.5$, $L=28$, $\beta t =40$, and $\sigma_0=e^2/2\pi$.} 
  \label{Cond.fig}
\end{figure}

The  enhancement of the velocity of the bipolarons  as a function of  phonon
frequency is also apparent in the optical conductivity, defined as
\begin{eqnarray}
  \sigma'(\omega) &&= 
  \frac{\pi}{Z \omega } \sum_{n,m} \rme^{-\beta E_m } (1-\rme^{-\beta \omega}) 
  \\\nonumber
  && \quad\times | \langle n | \hat{\jmath} | m \rangle|^2 
    \delta( E_n -  E_m - \omega )\,,
\end{eqnarray}
with the paramagnetic  current operator $\hat{\jmath} = \rmi t \sum_{i\sigma } ( 
\hat{c}^{\dag}_{i\sigma}  \hat{c}^\nag_{i + 1\sigma } - {\rm H.c.} ) $.
Results for $\sigma'(\omega)$ are plotted in figure~\ref{Cond.fig}, and
reveal that the Drude weight
is enhanced by an order of magnitude upon increasing the phonon frequency
from $\omega_0/t = 0.1$ to  $\omega_0/t = 0.5$.  

The enhancement of the Drude peak follows directly from the dynamical
charge structure factor.  Using the continuity equation, the  optical
conductivity and the dynamical charge structure factor are related via \cite{Assaad08}
\begin{equation}
  \label{SigmavsNq.eq}
  \sigma'( q,\omega)   = \frac{\omega}{ q^2 } 
  \left( 1 - \rme^{-\beta \omega} \right) N( q,\omega)\,.
\end{equation}
In the long wavelength limit, we can make the approximation $N(q,\omega)
\propto  q  \delta( \omega  - v_\mathrm{c} q )$. Here, $v_\mathrm{c}$ is the
charge velocity and the prefactor $q$ stems from  phase space available for
long-wavelength  charge fluctuations.\footnote{The factor $q$ equally
follows from the sum rule $\int\rmd \omega  N(q,\omega)  =
N(q) \propto q$}  This ansatz leads to  $\sigma'(q = 0,\omega)  \propto
v_\mathrm{c} \delta(\omega)$, so that the enhancement of the velocity
$v_\mathrm{c}$ is reflected in the conductivity. 

\section{Discussion and conclusions}\label{sec:discussion}

We have presented results for static and dynamical correlation functions of
the quarter-filled Holstein model at $\lambda = 0.35$, and concentrated  on
the evolution of the metallic state from the adiabatic to the anti-adiabatic
limit. For all considered values of the phonon frequency, our data support
the presence of a spin gap as a consequence of backward scattering. Such a
system with a gapless charge mode and a gapped spin mode is described
by the Luther-Emery fixed point. At the latter, the density and pairing
correlators have the form \cite{Giamarchi}
\begin{eqnarray} \label{LE.eq}
  \langle n(r) n(0) \rangle 
  &&= 
  \frac{A_0}{ r^2 } + \frac{A_1}{ r^{K_\rho} } \cos(2\kF r )\\
  &&\quad+  \frac{A_2}{r^{4 K_{\rho} } } \cos(4 \kF r )   
  + \dots\,,  
  \nonumber \\
  \langle \Delta^{\dag}(r) \Delta(0) \rangle 
  &&=  
  \frac{C} {r^{1/K_\rho}} + \dots\,,
\end{eqnarray}
in addition to an exponential decay of the spin-spin correlations.
The dominant correlations are $2\kF$ density correlations for $K_\rho
< 1$, and pairing correlations for $K_\rho  > 1 $. Our numerical results, in
particular those shown in figure~\ref{fig:cdw-pairing}, are hence best
understood in terms of an enhancement of $K_\rho$ from values
smaller than one at small $\omega_0/t$ (leading to dominant
charge-density-wave correlations) to values larger than $1$ in the
nonadiabatic regime where we observe significantly larger pairing
correlations. A quantitative finite-size scaling of $K_\rho$ is difficult
for several reasons, including the necessity to increase $L$ in steps
of eight, and the possible importance of logarithmic corrections as
a result of retardation effects \cite{PhysRevB.84.165123}. Nevertheless, a
cubic extrapolation based on $L=12,20,28$ gives $K_\rho\approx0.8$ for
$\omega_0/t=0.1$ and $K_\rho\approx1.1$ for $\omega_0/t=4$.

Since the spin degrees of freedom always remain gapped, we have interpreted
our results in terms of preformed pairs  which are essentially localised in
the $2\kF$ charge-density-wave state realised in the adiabatic limit.
This localisation leads to a  metallic state with a small
Drude weight. Upon increasing  the phonon frequency, enhanced lattice
fluctuations permit bipolaron motion, thereby providing  a  kinetic energy
gain. The resulting state is characterised by a significantly larger Drude
response  or, equivalently, an enhanced bipolaron velocity. 

A fruitful direction for future research concerning the evolution of
the Peierls state is the effect of higher dimensions. By studying, for
example, weakly coupled Holstein chains, the crossover from a
charge-density-wave state to a superfluid discussed here may evolve into a
phase transition expected to be in the XY universality class.

\ack

We acknowledge support from the DFG Grants No.~Ho~4489/2-1 and FOR 1162, and
generous computer time at the LRZ Munich and the J\"ulich Supercomputing Centre.


\end{document}